\documentclass[epj,nopacs]{svjour}

\usepackage{graphics}
\usepackage[usenames,dvipsnames]{color}
\usepackage{graphicx} 
\usepackage{dcolumn}
\usepackage{bm}
\usepackage{epsfig}
\usepackage{pdflscape}
\usepackage{subfigure} 
\usepackage{multirow}
\usepackage[colorlinks=true,citecolor=blue,filecolor=blue,linkcolor=blue,urlcolor=blue,pdftex]{hyperref}
\usepackage{rotating}
\usepackage{longtable}
\usepackage{lineno}

\begin{document}
\title{Lowering the radioactivity of the photomultiplier tubes for the XENON1T dark matter experiment}

\author{E.~Aprile\inst{1} 
\and F.~Agostini\inst{2,}\inst{16} 
\and M.~Alfonsi\inst{3} 
\and L.~Arazi\inst{4} 
\and K.~Arisaka\inst{5} 
\and F.~Arneodo\inst{6} 
\and M.~Auger\inst{7} 
\and C.~Balan\inst{8} 
\and P.~Barrow\inst{7} 
\and L.~Baudis\inst{7}\thanks{\emph{Email}: lbaudis@physik.uzh.ch}
\and B.~Bauermeister\inst{9} 
\and A.~Behrens\inst{7} 
\and P.~Beltrame\inst{4}\thanks{Present address: University of Edinburgh, UK}
\and A.~Brown\inst{10} 
\and E.~Brown\inst{11} 
\and S.~Bruenner\inst{12} 
\and G.~Bruno\inst{13} 
\and R.~Budnik\inst{4} 
\and L.~B\"utikofer\inst{14} 
\and J.~M.~R.~Cardoso\inst{8} 
\and D.~Coderre\inst{14} 
\and A.~P.~Colijn\inst{3} 
\and H.~Contreras\inst{1} 
\and J.~P.~Cussonneau\inst{15} 
\and M.~P.~Decowski\inst{3} 
\and A.~Di~Giovanni\inst{6} 
\and E.~Duchovni\inst{4} 
\and S.~Fattori\inst{9} 
\and A.~D.~Ferella\inst{16} 
\and A.~Fieguth\inst{13} 
\and W.~Fulgione\inst{16} 
\and M.~Garbini\inst{2} 
\and C.~Geis\inst{9} 
\and L.~W.~Goetzke\inst{1} 
\and C.~Grignon\inst{9} 
\and E.~Gross\inst{4} 
\and W.~Hampel\inst{12} 
\and R.~Itay\inst{4} 
\and F.~Kaether\inst{12} 
\and G.~Kessler\inst{7} 
\and A.~Kish\inst{7} 
\and H.~Landsman\inst{4} 
\and R.~F.~Lang\inst{10} 
\and M.~Le~Calloch\inst{15} 
\and D.~Lellouch\inst{4} 
\and L.~Levinson\inst{4} 
\and C.~Levy\inst{11} 
\and S.~Lindemann\inst{12} 
\and M.~Lindner\inst{12} 
\and J.~A.~M.~Lopes\inst{8}\thanks{Also at Coimbra engineering institute, Coimbra, Portugal}
\and A.~Lyashenko\inst{5} 
\and S.~Macmullin\inst{10} 
\and T.~Marrod\'an~Undagoitia\inst{12}\thanks{\emph{Email}: marrodan@mpi-hd.mpg.de}
\and J.~Masbou\inst{15} 
\and F.~V.~Massoli\inst{2} 
\and D.~Mayani\inst{7} 
\and A.~J.~Melgarejo~Fernandez\inst{1} 
\and Y.~Meng\inst{5} 
\and M.~Messina\inst{1} 
\and B.~Miguez\inst{17} 
\and A.~Molinario\inst{17} 
\and G.~Morana\inst{2} 
\and M.~Murra\inst{13} 
\and J.~Naganoma\inst{18} 
\and U.~Oberlack\inst{9} 
\and S.~E.~A.~Orrigo\inst{8} \thanks{Present address: IFIC, CSIC-Universidad de Valencia, Spain}
\and P.~Pakarha\inst{7} 
\and E.~Pantic\inst{5} 
\and R.~Persiani\inst{2} 
\and F.~Piastra\inst{7} 
\and J.~Pienaar\inst{10} 
\and G.~Plante\inst{1} 
\and N.~Priel\inst{4} 
\and L.~Rauch\inst{12} 
\and S.~Reichard\inst{10} 
\and C.~Reuter\inst{10} 
\and A.~Rizzo\inst{1} 
\and S.~Rosendahl\inst{13} 
\and J.~M.~F.~dos Santos\inst{8} 
\and G.~Sartorelli\inst{2} 
\and S.~Schindler\inst{9} 
\and J.~Schreiner\inst{12} 
\and M.~Schumann\inst{14} 
\and L.~Scotto~Lavina\inst{15} 
\and M.~Selvi\inst{2} 
\and P.~Shagin\inst{18} 
\and H.~Simgen\inst{12} 
\and A.~Teymourian\inst{5} 
\and D.~Thers\inst{15} 
\and A.~Tiseni\inst{3} 
\and G.~Trinchero\inst{17} 
\and C.~Tunnell\inst{3} 
\and O.~Vitells\inst{4} 
\and R.~Wall\inst{18} 
\and H.~Wang\inst{5} 
\and M.~Weber\inst{12} 
\and C.~Weinheimer\inst{13} (The XENON Collaboration)
\and  M.~Laubenstein\inst{16} 
}                    
\institute{Physics Department, Columbia University, New York, NY, USA  
\and 
Department of Physics and Astrophysics, University of Bologna and INFN-Bologna, Bologna, Italy
\and
Nikhef and the University of Amsterdam, Science Park, Amsterdam, Netherlands
\and
Department of Particle Physics and Astrophysics, Weizmann Institute of Science, Rehovot, Israel
\and
Physics \& Astronomy Department, University of California, Los Angeles, CA, USA
\and
New York University Abu Dhabi, Abu Dhabi, United Arab Emirates
\and
Physik-Institut, University of Zurich, Zurich, Switzerland
\and
Department of Physics, University of Coimbra, Coimbra, Portugal
\and
Institut f\"ur Physik \& Exzellenzcluster PRISMA, Johannes Gutenberg-Universit\"at Mainz, Mainz, Germany
\and
Department of Physics and Astronomy, Purdue University, West Lafayette, IN, USA
\and
Department of Physics, Applied Physics and Astronomy, Rensselaer Polytechnic Institute, Troy, NY, USA
\and
Max-Planck-Institut f\"ur Kernphysik, Heidelberg, Germany
\and
Institut f\"ur Kernphysik, Wilhelms-Universit\"at M\"unster, M\"unster, Germany
\and
Albert Einstein Center for Fundamental Physics, University of Bern, Bern, Switzerland
\and
Subatech, Ecole des Mines de Nantes, CNRS/In2p3, Universit\'e de Nantes, Nantes, France
\and
INFN-Laboratori Nazionali del Gran Sasso and Gran Sasso Science Institute, L'Aquila, Italy
\and
INFN-Torino and Osservatorio Astrofisico di Torino, Torino, Italy
\and
Department of Physics and Astronomy, Rice University, Houston, TX, USA
}

\date{Received: date / Revised version: date}
\abstract{
The low-background, VUV-sensitive 3-inch diameter photomultiplier tube R11410 has been developed by Hamamatsu for dark matter direct detection experiments using liquid xenon as the target material. We present the results from the joint effort between the XENON collaboration and the Hamamatsu company to produce a highly radio-pure photosensor (version R11410-21) for the XENON1T dark matter experiment.
After introducing the photosensor and its components, we show the methods and results of the radioactive contamination measurements of the individual materials employed in the photomultiplier production.  We then discuss the adopted strategies to reduce the radioactivity of the various PMT versions. Finally, we detail the results from screening 216 tubes with ultra-low background germanium detectors,  as well as their implications for the expected electronic and nuclear recoil background of the XENON1T experiment.
\PACS{
      {PACS-key}{discribing text of that key}   \and
      {PACS-key}{discribing text of that key}
     } 
} 
\authorrunning{}
\titlerunning{Lowering the radioactivity of the photomultiplier for the XENON1T dark matter experiment}
\maketitle

\section{Introduction}
\label{sec:intro}

Among the various experimental methods for the direct detection of dark matter particles,  liquid-xenon (LXe) time projection chambers have demonstrated the highest sensitivities over the past years\,\cite{Baudis:2012ig,Baudis:2014naa}. The XENON100 experiment\,\cite{Aprile:2011dd} did not find any evidence for dark matter and published the world's best upper limits on  spin-indepen\-dent\,\cite{Aprile:2012nq} and spin-dependent\,\cite{Aprile:2013doa} couplings of  weakly interacting massive particles (WIMPs)  to nucleons and neutrons, respectively.  Recently, the LUX experiment  has confirmed and improved upon these results, reaching an upper limit on the spin-independent WIMP-nucleon cross section of $7.6\times10^{-46}$\,cm$^2$  at a WIMP mass of 33\,GeV/$c^2$\,\cite{Akerib:2013tjd}. 

To significantly  increase the current experimental sensitivities, the XENON collaboration builds the XENON1T experiment\,\cite{Aprile:2012zx}. The construction work 
at the Laboratori Nazionali del Gran Sasso (LNGS) in Italy started in autumn 2013 and will continue until summer 2015. Detector commissioning, and a first science run are expected for mid and late 2015, respectively, while the design dark matter sensitivity will be reached after two years of operation. With 35 times more target mass, and a background goal 100 times lower than that of XENON100\,\cite{Aprile:2011vb,Aprile:2013tov},  the XENON1T sensitivity to spin-independent WIMP-nucleon cross sections is expected to be $2\times10^{-47}$\,cm$^2$ at a WIMP mass around 40\,GeV/$c^2$. This ambitious goal demands an ultra-low background in the central detector region, implying a very low radioactivity of all detector construction materials, in particular  those in the vicinity of the xenon target. Among these, the photosensors, that detect the primary and secondary scintillation light produced in a WIMP-nucleus collision, are the most complex, multi-material components.  Photomultiplier tubes (PMTs) were shown to cause a major contribution to the electronic recoil background from detector materials in the XENON100\,\cite{Aprile:2011vb} and LUX\,\cite{Akerib:2012ys} experiments. Other current and planned experiments based on xenon as a detection medium for rare-event searches are XMASS\,\cite{ref::xmass}, PandaX\,\cite{Cao:2014jsa}, LZ\,\cite{Malling:2011va}, DARWIN\,\cite{Baudis:2012bc}, NEXT\,\cite{Lorca:2012dv} and EXO\,\cite{Auger:2012gs}. 

Many of these projects employ PMTs for signal readout, therefore the availability of ultra-low background,  high-quantum efficiency sensors, and capable to operate stably for long periods at low temperatures  is of crucial importance.  PandaX, XENON1T, LZ and possibly DARWIN use or plan to use a new, 3-inch diameter tube recently developed by Hamamatsu\,\cite{Hamamatsu}. The tube was extensively tested at room temperature and in liquid xenon\,\cite{Lung:2012pi,Baudis:2013xva}, and first measurements of the radioactivity of an early PMT version were presented in\,\cite{Baudis:2013xva,Aprile:2011ru,Akerib:2012da}.

In this paper we present the most recent version, named R11410-21, which has been developed by Hamamatsu together with the XENON collaboration to produce a tube that meets the strict background requirements of the XE\-NON1T experiment.  In Section\,\ref{sec:sensor} we summarise the general characteristics of the PMT. In Section\,\ref{sec:techniques} we introduce the measurement techniques that were employed to evaluate the radioactive contamination of individual PMT components and of the PMT as a whole. In Section\,\ref{sec:results}, we present the main screening results for the construction materials and the first PMT production batches. Finally, in Section\,\ref{sec:impact} we interpret the results and discuss their impact on the expected backgrounds of the XENON1T experiment.

\section{The R11410 photosensor}
\label{sec:sensor}

The R11410 photomultiplier is a 3-inch diameter tube produced by Hamamatsu for xenon-based dark matter detectors. It operates stably at typical temperatures and pressures in a LXe detector, around $-100^{\circ}$C and 2\,atm, respectively.  Apart from a greatly reduced intrinsic radioactivity level, as we will show in the following sections, a major advantage is its high quantum efficiency (QE)  at  the xenon scintillation wavelength of 175\,nm. A mean value of $\langle\textrm{QE}\rangle\, = 35$\% has been achieved for the tubes delivered for XENON1T and a few tubes have a QE as high as 40\%. Along with 90\% electron collection efficiency\,\cite{Lung:2012pi}, the tube ensures a high detection efficiency for VUV scintillation photons produced by particle interactions in xenon.

\begin{figure}[h]
\includegraphics[width=0.5\textwidth]{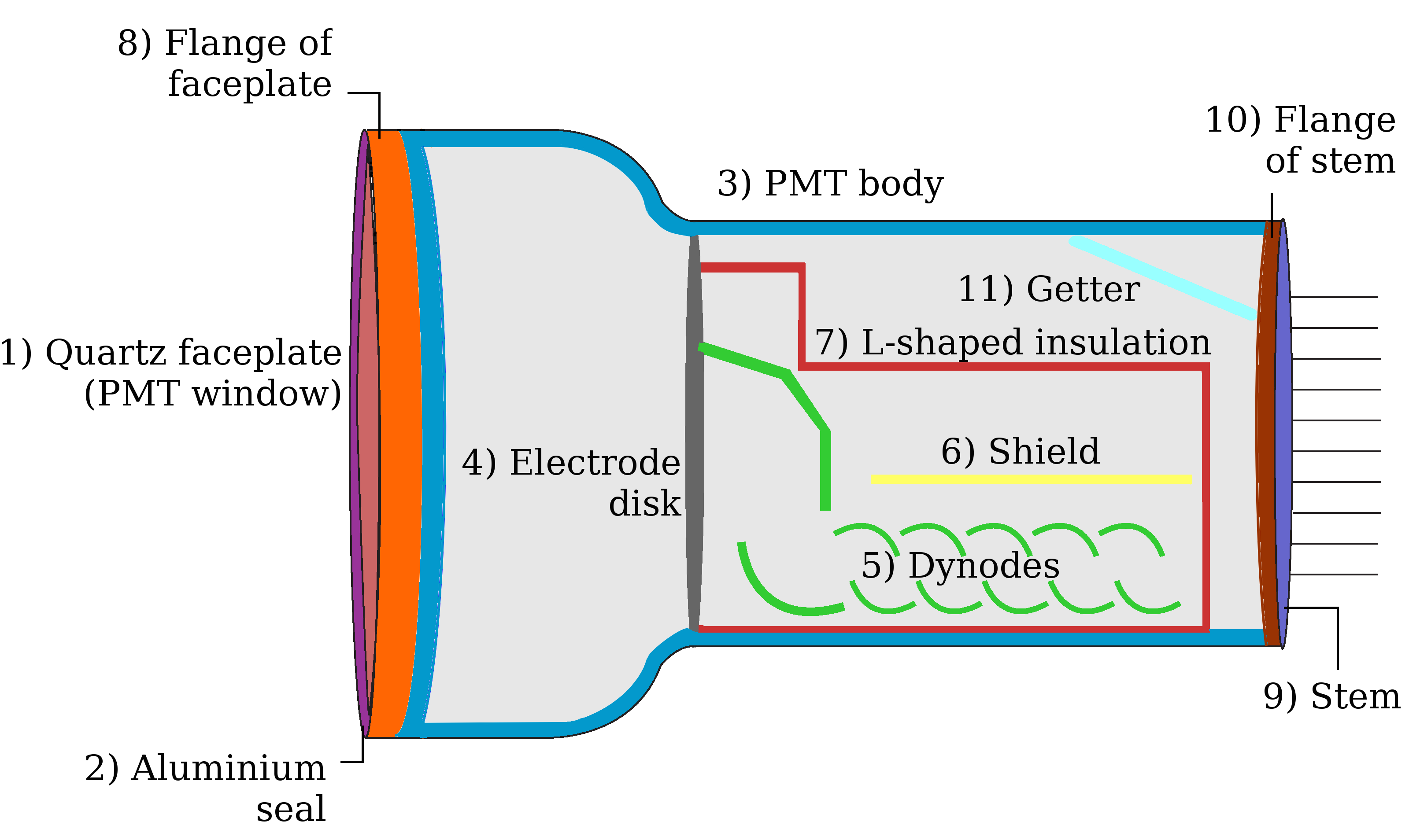} 
\caption{A schematic figure of the R11410 PMT, showing its main components. The numbers correspond to those in Table\,\ref{tab:samples}, the colours to those in Figure \ref{fig::bar_plot}.} 
\label{fig:pmt_schematic}
\end{figure}

The R11410 photomultiplier  has a  VUV-transparent quartz window and a low-temperature bialkali photocathode deposited on it. A 12-dynode electron-multiplication system provides a signal amplification of about $3.5\times10^6$ at $-1500$\,V operating voltage. The peak-to-valley ratio is at least 2, showing a good separation of the single photoelectron signal from the noise spectrum\,\cite{Baudis:2013xva}. Inside the tube, the dynodes are insulated using  L-shaped quartz plates. The body of the PMT is about 115\,mm long\,\cite{Datasheet_R11410}, and it is made out of a Kovar alloy, most of which has a very low cobalt content (``cobalt-free Kovar"). On the back side of the PMT, the stem uses ceramic material to insulate the connections to the individual dynodes. The tube has been intensively tested to demonstrate its stable behaviour during long-term operation in LXe and in the presence of an electric field\,\cite{Baudis:2013xva}. At $-100^{\circ}$C, this PMT shows a low dark count rate of about 80\,Hz at $-1600$\,V above an area of the single photoelectron (PE) peak of 1/3\,PE.
The corresponding single photoelectron resolution is between $(35-40)$\%\,\cite{Baudis:2013xva}. In Figure \ref{fig:pmt_schematic} we show a sketch of the R11410  PMT, displaying its main components that were introduced in this section.

\section{Techniques and measurements}
\label{sec:techniques}

We now briefly describe the techniques used to measure the radioactive contamination of the PMTs and of the various PMT components. The most relevant isotopes are those of the natural uranium and thorium decay chains. In addition to the gamma rays, the emitted $\alpha$-particles can produce fast neutrons via $(\alpha,n)$ reactions in a given material.  These neutrons can cause nuclear recoils in the LXe, similar to the expected interactions of WIMPs, and thus present a serious source of background. We also study the $^{60}$Co, $^{40}$K and $^{137}$Cs isotopes, among others, because their gamma-emission contributes to the electron-recoil background of the experiment.  This contribution can, however, be reduced by signal/background discrimination in xenon time projection chambers that measure both the light and charge signals produced in a particle interaction in the active detector volume.
 
\subsection{Germanium spectroscopy}

Gamma-ray spectroscopy is a standard method to  screen and select materials for rare-event searches. Germanium spectrometers in dedicated low-background shields combine a high energy resolution with a very low intrinsic background.  We have employed the world's most sensitive high-purity germanium (HPGe) spectrometers, the GeMPI\,\cite{Neder:2000,Heusser:2004,Budjas:2007yj} and Gator\,\cite{Baudis:2011am} detectors. They are located at LNGS, under an average overburden of 3600\,m\,w.\,e., where the muon flux is reduced to $\sim$1\,m$^{-2}$\,h$^{-1}$. The detectors are coaxial, p-type HPGe crystals, with masses around $(2.2\,-\,2.3)$\,kg, housed in electro-refined copper cryo\-stats. The cryostats are surrounded by shields made of low-activity copper, lead and polyethylene, and are housed in air-tight boxes which are permanently flushed with pure nitrogen gas to suppress the influx of $^{222}$Rn. The sample chambers have dimensions of  $\sim (25\times 25\times 30)$\,cm$^3$, allowing the placement of rather large samples around the germanium detectors. Sample handling chambers equipped with airlocks and glove boxes are located above the shields, with compartments that allow for the storage of samples prior to their measurement and thus for the decay of $^{222}$Rn, $^{220}$Rn and their progenies. Before introducing the material samples into the airlock, their surface is cleaned. The samples are degreased with an acid soap and rinsed with deionized water in an ultrasonic bath. Finally, the surfaces are cleaned with ethanol.   The PMTs are also cleaned before the measurements, their surface is wiped with ethanol.

Sealed calibration sources can be fed into the inner chambers  in order to calibrate the energy scale and to determine the energy resolution. The relative efficiency of the Ge spectrometers is around 100\%. In $\gamma$-spectroscopy, the quoted efficiency is defined relative to a $7.62\,\O \times7.62$\,cm NaI(Tl) crystal, for the 1.33\,MeV $^{60}$Co photo-absorption peak at a source-detector distance of 25\,cm\,\cite{Knoll:2000fj}. The integral counting rate in the energy region $(100-2700)$\,keV  is in the range $(6-10)$\,events/h.  They can detect sample activities of the order  $\sim\,10\,\mu$Bq/kg for  $^{226}$Ra and $^{228}$Th/$^{228}$Ra. To determine the concentration of a radionuclide in a given sample, the most prominent gamma lines in the spectrum are analysed, after subtraction of the background spectrum. 
The detection efficiencies for each line are determined in Monte Carlo simulations which are run individually for each type of sample and the full detector-sample geometry. The screening time is typically $2-3$\,weeks for each of these results. Upper limits are calculated as described in~\cite{Hurtgen:2000} and given at 90\%\,C.L. More details about the analysis procedure can be found in~\cite{Baudis:2011am}.

\subsection{Glow-discharge mass  spectrometry}

To detect trace impurities from solid samples, glow-dis\-charge mass  spectrometry (GDMS) was used. The measurements presented here were carried out by the Evans Analytical Group (EAG)\,\cite{Evans} on behalf of the XENON collaboration. The samples are exposed to a gas discharge or plasma atomisation/ionisation source. Argon is used as discharge gas to sputter the atoms out from the sample. The ion beam is then directed into a system of focusing lenses and cones, which finally inject the ions into a quadrupole mass spectrometer. The uncertainty of the result is between $(20-30)$\%. The detection limit is better than $10^{-9}$\,g/g (ppb) for conductors and an order of magnitude higher for non-conductors. This method detects traces of uranium and thorium in a sample, which are then translated into contaminations on $^{238}$U and $^{232}$Th while germanium counting mainly measures the $^{226}$Ra and $^{228}$Th/$^{228}$Ra isotopes of the U and Th chains in a given material.  Assuming secular equilibrium in the decay series, the conversion factors from units of Bq/kg to concentrations by weight, which are more common in mass spectrometry, are\,\cite{Heusser:1995wd}: 1\,Bq $^{238}$U/kg\,$\mathrel{\widehat{=}}$\,81$\times$10$^{-9}$g\,U/g (81\,ppb U) and 1\,Bq $^{232}$Th/kg\-\,$\mathrel{\widehat{=}}$\,246$\times$10$^{-9}$g\,Th/g (246\,ppb Th). The GDMS measurement is particularly relevant for the early part of the $^{238}$U chain, which can be barely detected via germanium spectroscopy due to the low energy and low intensity of the emitted gamma lines.

\subsection{Overview of the PMT screening measurements}

The radioactivity of the materials employed to manufacture the R11410 tube has been gradually reduced from one version to the next, with the goal of minimizing the overall radioactivity of the final product. The first R11410 version used a glass stem, standard Kovar alloy for the body, and ceramic as a dynode insulator. The seal between the body and the quartz window was made with aluminium of standard (std.) purity.  As this tube version still had a fairly high radioactivity level\,\cite{Aprile:2011ru},  the subsequent versions were targeted at replacing the main components with new materials such as Co-free Kovar alloy for the body of the tube, quartz instead of ceramic as an insulator, and high-purity aluminium for the seal.  We note that the second PMT version was originally delivered as R11410-MOD and later renamed to R11410-10. The R11410-21 version employs low-background materials selected by us (see Section \ref{sec:samples}) for those PMTs to be operated in XENON1T only. In an attempt to further reduce the radioactivity of the tube, a stem made out of sapphire has been proposed by Hamamatsu, R11410-30.  It will not be employed in XENON1T, for the overall background reduction would be insignificant, as we detail in  Section \ref{sec:impact}.
The modifications that occurred between the different PMT versions are summarised in Table\,\ref{tab:PMTversMat}. 

\begin{table*}
\caption{\small Main material components used across the different R11410 PMT versions. $^{(1)}$For the production of  the R11410-21 tube, that features the same design as the R11410-20 version, very low-background materials were selected to fulfil the demanding background requirements of the XENON1T experiment.} \label{tab:PMTversMat}
\begin{center}
\begin{tabular}{llllll}
	\hline
Part &{ R11410} & { R11410-10} & { R11410-20}  & { R11410-21$^{(1)}$} & { R11410-30} \\
		\hline
 Stem &   Glass & Ceramic & Ceramic & Ceramic & Sapphire \\
 Body &   Kovar alloy & Kovar alloy & Co-free Kovar & Co-free Kovar& Co-free Kovar \\
Insulator &   Ceramic & Quartz & Quartz & Quartz & Quartz \\
 Al seal &   Std. purity & Std. purity  & High purity  & High purity  & High purity  \\
	\hline
\end{tabular}
\end{center}
\end{table*}

\section{Results}\label{sec:results}

In this section, we first present the measurements of the intrinsic radioactive contamination of the individual phototube construction materials.  These are followed by the germanium screening results for the fully operational PMTs, and a comparison with other PMT types operated in rare event searches based on liquid xenon.

\subsection{Material samples}
\label{sec:samples}

A total of 12 material samples,  most of these to be used for the R11410-21 PMT construction, have been measured. Table\,\ref{tab:samples} compiles a list of the studied materials, the sample masses used for germanium screening, as well as the mass of each material employed for manufacturing  of one photomultiplier tube. 
 \begin{table*}
 \caption{List of  measured material samples including the mass of each component per PMT.   $^{(1)}$The sapphire (sample~12) is an alternative material to the ceramic (sample~9) for the stem  of a possible future PMT version. }\label{tab:samples}
 \begin{center}
\begin{tabular}{ c lcc}
	\hline
{ Nr. } &{ Sample} &{Mass for Ge-screening  [g] } & {Mass per PMT  [g]}  \\
	 	\hline
1 &Quartz: faceplate (PMT window) &1179& 30  \\
2&Aluminium: pure Al for sealing &515& 0.6   \\
3&Kovar: Co-free body &500& 78 \\
4&Stainless steel sheets: electrode disk  &555 & 8.2   \\ 
5&Stainless steel sheets: dynodes   &510 & 7.2   \\
6&Stainless steel: shield & 519 &4  \\
7& Quartz: L-shaped insulation &838 & 14.4     \\
8&Kovar: flange of faceplate &525 & 18   \\
9a&Ceramic: stem & 597 & 16   \\
9b&Ceramic: stem & 498 & 16  \\
10&Kovar: flange of ceramic stem &511  & 14    \\
11 & Getter (10 pieces) & 0.58 & 5.8$\times$10$^{-2}$    \\
12&Sapphire (for R11410-30)$^{(1)}$ & 243 & 16  \\
	\hline
\end{tabular}
\end{center}
\end{table*}
For most components, the mass of the screened material was around or above 500\,g, with exception of the quartz, sapphire and getter samples.  For the analysis with the GDMS technique, typically $\mathcal{O}(10)$\,g of  each material were needed.    
The results of the germanium and mass spectrometry measurements for the isotopes of the uranium and thorium chains are summarised in Table\,\ref{tab:results_U_Th}. Upper limits are given at 90\% confidence level, and the quoted errors include statistical and systematic errors.   The item numbers correspond to the materials as listed in Table\,\ref{tab:samples}.  
 
 \begin{table*}
\caption{Results of the germanium (Ge) screening and the mass  spectroscopy (GDMS) in mBq/PMT for isotopes in the $^{238}$U, $^{235}$U and $^{232}$Th chains.  The sample numbers correspond to those in Figure \ref{fig:pmt_schematic}, Table \ref{tab:samples} and Figure \ref{fig::bar_plot}. For the GDMS measurement, only results for $^{238}$U and $^{232}$Th are available. The ceramic material (sample 9) was measured twice, with different HPGe detectors and analyzed by different groups, as a cross check of the sensitivity. The  {\sl total upper limit (L) } is calculated using the measured contamination level or the inferred upper limit  while the {\sl total detection (D)} considers only detected values. Sample 12 (alternative material to ceramic) in not considered in the sums listed in the table. }\label{tab:results_U_Th}
\begin{center}
\small\addtolength{\tabcolsep}{-5pt}
\begin{tabular}{ c c ccccc}
	\hline
{ Sample } &{ Method} & {$^{238}$U} & {$^{226}$Ra}  &  {$^{228}$Ra} ({ $^{232}$Th}) \ \ \ &  {$^{228}$Th}  &  {$^{235}$U }\\
{ nr.}         & { (detector)} &                    &                        &                                                       &                         &                       \\
	 \hline
1 & Ge & $<0.33$ & \ $3.6(6)\cdot10^{-2}$ \  & $<1.2\cdot10^{-2}$ & $<1.1\cdot10^{-2}$ & $<1.1\cdot10^{-2}$ \\
1 & GDMS & $<1.8$& -- &  $<0.63$ &-- & -- \\
2 & Ge & \ $<2.8\cdot10^{-2}$ \ \  & $<7.2\cdot10^{-4}$ & $<6.6\cdot10^{-4}$ & \ $6.6(3)\cdot10^{-4}$ \  & $<5.5\cdot10^{-4}$ \\
2 & GDMS & $<0.15\cdot10^{-2}$& -- & $<2.5\cdot10^{-4}$ &-- & -- \\
3 & Ge& $<9.2$ & $<0.26$ & $<0.36$ & $<0.34$ & $<0.17$ \\
3 & GDMS& $<9.5\cdot10^{-2}$& -- & $<3.2\cdot10^{-3}$ & -- & -- \\
4 & Ge& $<0.90$ & $<2.5\cdot10^{-2}$ & $<4.3\cdot10^{-2}$ & $<3.3\cdot10^{-2}$ & $<1.9\cdot10^{-2}$ \\
4 & GDMS& $5.0\cdot10^{-2}$& -- & $0.68\cdot10^{-2}$ & --& -- \\
5 & Ge & $<0.53$ & $2.7(6)\cdot10^{-2}$ & $<9.4\cdot10^{-3}$ & $<9.4\cdot10^{-3}$ & $<1.1\cdot10^{-2}$ \\
5 & GDMS& $<0.17$& -- & $<15\cdot10^{-3}$ & -- & -- \\
6 & Ge& $<0.37$ & $1.3(2)\cdot10^{-2}$ & $8(4)\cdot10^{-3}$ & $8(4)\cdot10^{-3}$ & $<5.2\cdot10^{-2}$ \\
6 & GDMS& $<3.4\cdot10^{-2}$& -- & $<17\cdot10^{-3}$ & -- & -- \\
7 & Ge & $<0.2$ & $2.9(3)\cdot10^{-2}$ & $<1.1\cdot10^{-2}$ & $<0.7\cdot10^{-2}$ & $<0.6\cdot10^{-2}$ \\
7 & GDMS& $<0.88$& -- & $<0.3$  & --& -- \\
8 & Ge& $<0.79$ & $3.7(7)\cdot10^{-2}$ & $<1.8\cdot10^{-2}$ & $<1.8\cdot10^{-2}$ & $<1.8\cdot10^{-2}$ \\
8 & GDMS& $4.4\cdot10^{-2}$ & -- & $0.98\cdot10^{-2}$& -- & -- \\
9 & Ge & $2.4(4)$ & $0.26(2)$ & $0.23(3)$ & $0.11(2)$ & $0.11(2)$ \\
9 & Ge & $3(1)$ & $0.30(3)$ & $0.21(3)$ & $0.11(2)$ & $0.11(3)$ \\
10 & Ge & $<0.65$ & $<8.3\cdot10^{-3}$ &  $<3\cdot10^{-2}$ &  $<7.5\cdot10^{-3}$ & $<1.7\cdot10^{-2}$ \\
10 & GDMS& $<1.7\cdot10^{-2}$ & -- & $<6\cdot10^{-3}$ & -- & -- \\
11 & Ge & $<0.7$ & $0.035(4)$ &  $<2\cdot10^{-2}$ &  $<2.9\cdot10^{-2}$  & $0.018(3)$ \\
	\hline
Total  L & Ge& $<16$ &  $<0.75$ &  $<0.74$&  $<0.67$ &  $<0.44$ \\
Total D &  Ge& $2.7$ &  $0.46$ &  $0.24$&  $0.12$ &  $0.13$ \\
Total L & GDMS& $<3.1$ & -- & $<1$ & -- & -- \\
Total D & GDMS& $0.09$ & -- & $0.02$ & -- & -- \\
	\hline
	Sample 12 & Ge & $<0.72$ & $0.72(4)$ &  $0.13(3)$ &  $0.08(2)$ & \ $<3.4\cdot10^{-2}$\  \ \\
	\hline
\end{tabular}
\end{center}
\end{table*}
  We remark that the presented results on $^{238}$U are obtained by inspecting the 92.4\,keV and 92.8\,keV lines from $^{234}$Th and the 1\,MeV line from $^{234m}$Pa. However, one can also infer the  $^{238}$U activity from the measured  activity for $^{235}$U, as we can assume their natural abundance (where the activities are related as A$_{235}\sim0.05\,\cdot $\,A$_{238}$). In general, the analysis $^{235}$U lines provide a higher sensitivity.  Secular equilibrium between various parts of the natural decay chains  may be broken  in processed materials, as the half-lives of $^{238}$U and $^{228}$Th and of some of their daughters are long, and in most cases equilibrium was not re-established. Nonetheless, the results of both techniques, HPGe screening and GDMS,  are compatible with one another. For the ceramic and sapphire samples only germanium screening results are reported due to large systematic uncertainties of the GDMS method for this type of material.
  The observed contamination with $^{40}$K,  $^{60}$Co and  $^{137}$Cs of each sample is shown in Table\,\ref{tab:results_k_co_cs}.  

\begin{table*}
\caption{\small Results of the germanium screening, in mBq/PMT, for $^{40}$K, $^{60}$Co and $^{137}$Cs for the main raw materials of the R11410-21 tube.  The totals L and D are calculated as in the previous table.} 
\label{tab:results_k_co_cs}
\begin{center}
\small\addtolength{\tabcolsep}{-5pt}
\begin{tabular}{c ccc }
	\hline
{ Sample nr.}  &  {$^{40}$K} &  {$^{60}$Co} &  {$^{137}$Cs }\\
		\hline
1  & $<8.1\cdot10^{-2}$ & $<4.5\cdot10^{-3}$ & $<4.8\cdot10^{-3}$\\
2   & $<5.7\cdot10^{-3}$ & $<1.2\cdot10^{-4}$ & $<2.3\cdot10^{-4}$\\
3  & $<0.99$ & $7(2)\cdot10^{-2}$ & $<0.10$\\
4  & $<6.4\cdot10^{-2}$ & $7.2(6)\cdot10^{-2}$ & $<8.0\cdot10^{-3}$\\
5  & $6(3)\cdot10^{-2}$ & $6(3)\cdot10^{-3}$ & $<7.9\cdot10^{-3}$\\
6  & \ \ $<3.2\cdot10^{-2}$ \ \ & \ \   $2(1)\cdot10^{-3}$ \ \  & \ \  $<2.6\cdot10^{-3}$\ \ \\
7  &  $7(3)\cdot10^{-2}$ & $<2.3\cdot10^{-3}$ & $<1.6\cdot10^{-3}$\\
8  & $7(4)\cdot10^{-2}$ & $0.26(2)$  & $9(4)\cdot10^{-3}$\\
9  & $1.1(2)$ & $<2\cdot10^{-2}$ & $<2\cdot10^{-2}$\\
9  & $1.6(2)$ & $<1.6\cdot10^{-2}$ & $<1.2\cdot10^{-2}$\\
10  & $7(4)\cdot10^{-2}$ & $0.22(2)$ &  $1.3(6)\cdot10^{-2}$\\
11  &  $8(3)\cdot10^{-2}$ & $<4.2\cdot10^{-3}$  &  $<4.2\cdot10^{-3}$ \\
	\hline
Total L &  {\bf $<3.1$} &  $<0.66$ & {\bf $<0.16$} \\
Total D &  {\bf $1.9$} &  $0.61$ & {\bf $0.021$} \\
	\hline
	Sample 12  & $0.14(6)$ & $<8.8\cdot10^{-3}$ &  $<1.5\cdot10^{-2}$\\
	\hline
\end{tabular}
\end{center}
\end{table*}
In addition to the 10 getter pieces, a getter extracted directly from a PMT was measured. As only one piece was available, the obtained upper limits for most isotopes are much above the sensitivity of those for other materials. Nonetheless, a contamination with $^{40}$K clearly above the one measured for the 10 getter samples is observed. Further investigations are ongoing. The discrepancy is likely due to processes that occur during the PMT assembly, implications are discussed in Section~\ref{sec:impact}. 

\begin{figure*}
\begin{center}
\includegraphics[width=0.8\textwidth]{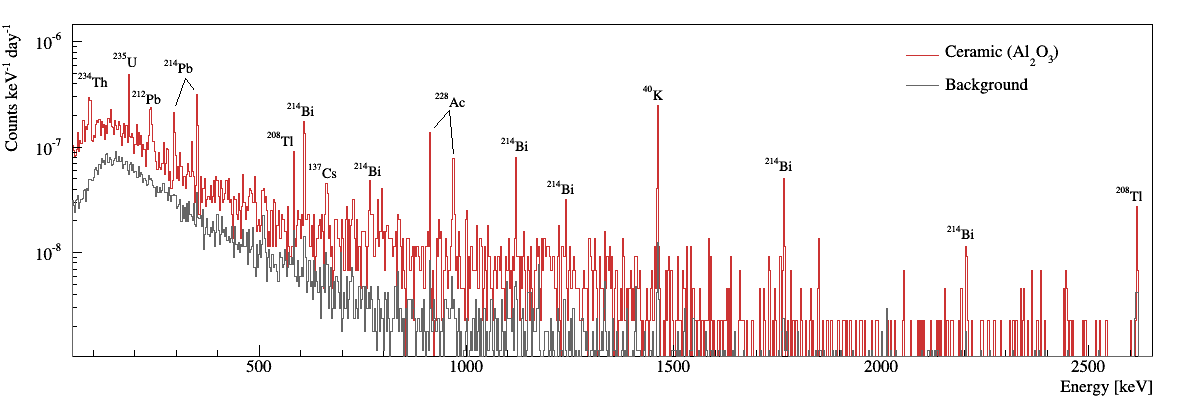} 
\includegraphics[width=0.8\textwidth]{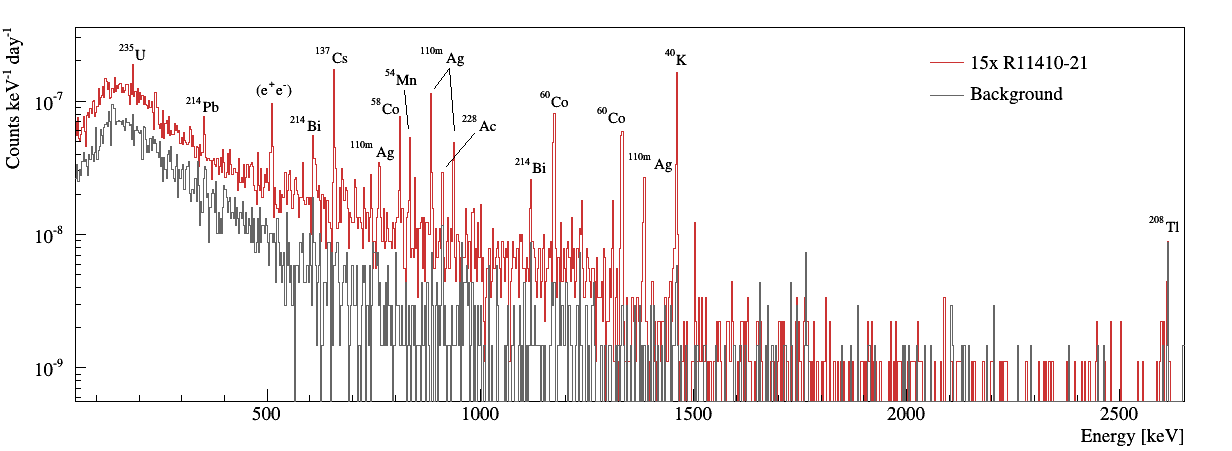} 
\end{center}
\caption{\small (Top) Spectrum of the ceramic sample (nr.\,9, red) along with the detector background (black). (Bottom) Spectrum of 15 Hamamatsu R11410-21 PMTs measured in July 2013 (red).  The background spectrum of the Gator detector together with the PTFE support structure for the PMTs is also shown (black).\ \label{fig:pmt_spectra}}
\end{figure*}

The sum of the measured contamination of all PMT parts, for a given isotope, is shown at the bottom of each table. We calculate a total upper limit (L) using either the measured contamination level, if available, or the upper limit from the most sensitive method, and a total detection (D) using only measured contaminations. These sums determine the range for the total expected radioactivity budget of a PMT, to be compared with the screening of the whole PMT, as presented in the next section.  In Figure \ref{fig:pmt_spectra} (top), we show as an example the spectrum of the ceramic sample, together with the background spectrum of the HPGe detector.
From all screened materials, the ceramic has the highest contribution to the total radioactivity of the tube, in particular regarding the  $^{238}$U component, which is at the level of 2.5\,mBq/PMT.

 \subsection{Measurements of PMT batches}
 
\begin{figure}
\begin{center}
\includegraphics[width=0.32\textwidth]{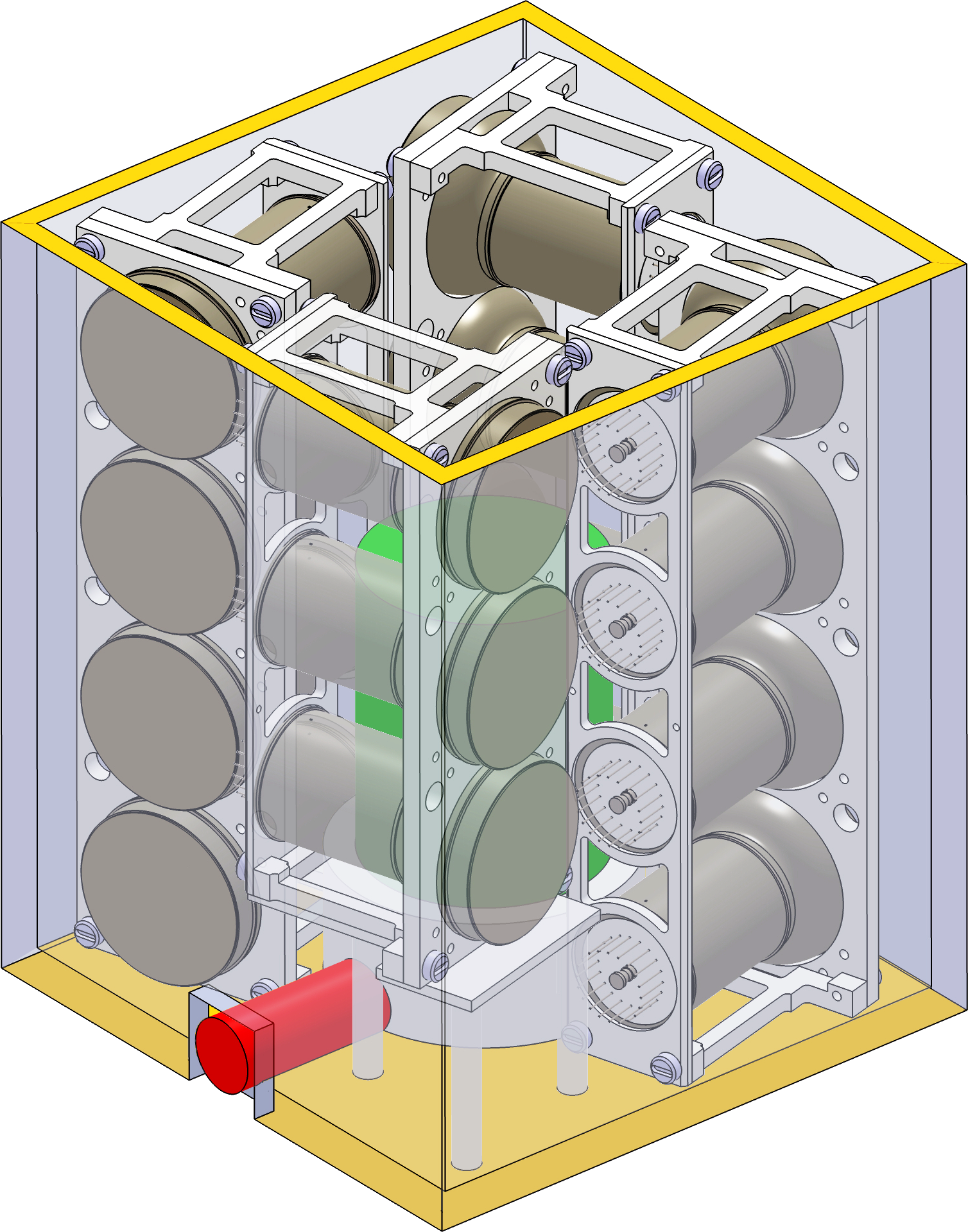} 
\end{center}
\caption{Schematic view of 15 R11410-21 PMTs in the Gator detector chamber, held by custom-made,  low-background PTFE support structure.  The germanium detector in its cryostat is seen in green in the centre. \label{fig:pmt_holder}}
\end{figure}

\begin{figure}
\includegraphics[width=0.5\textwidth]{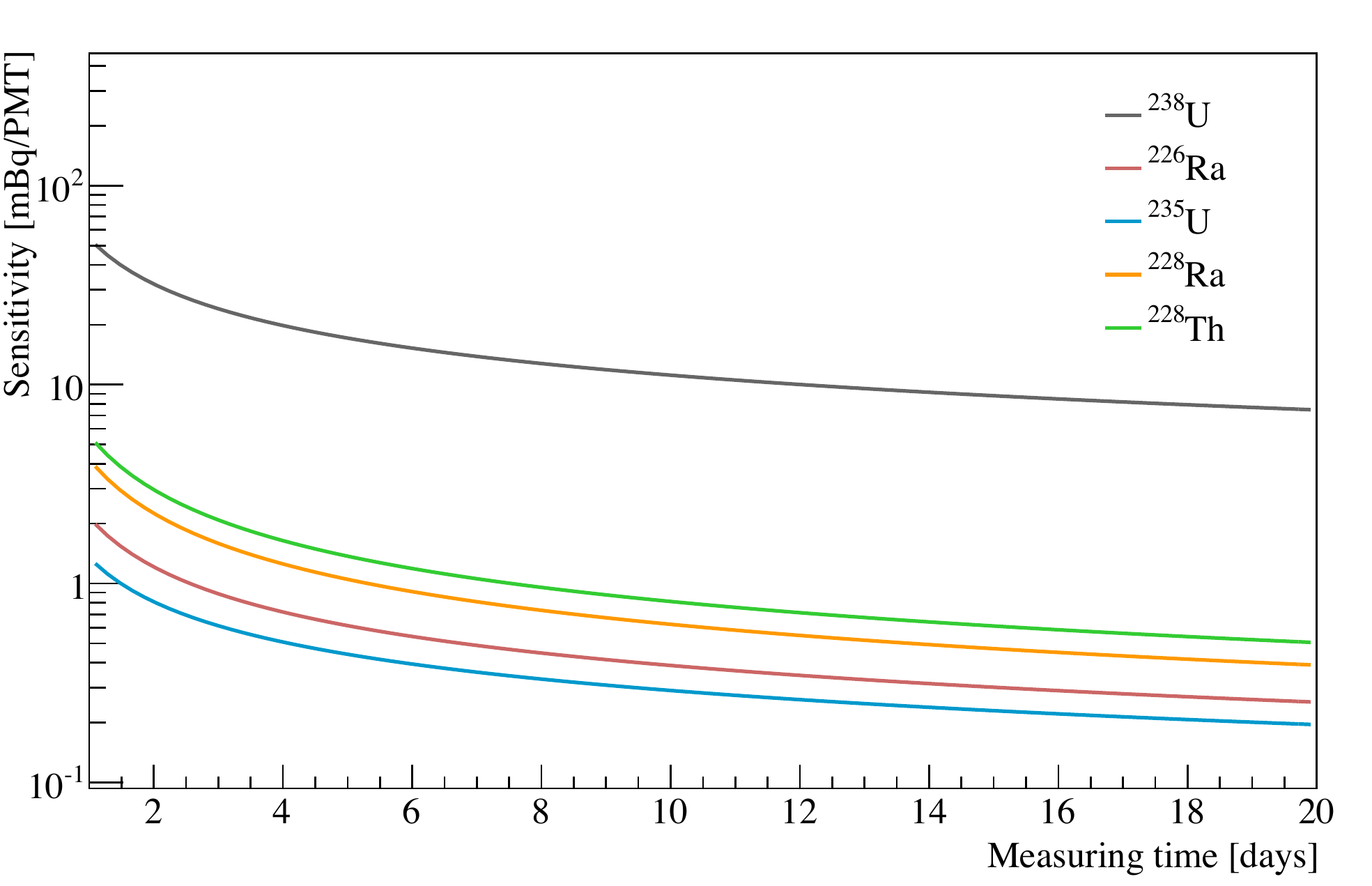} 
\caption{Sensitivity for various isotopes as a function of measuring time, assuming  that 15 PMTs are placed inside the detector chamber. 
\label{fig:pmt_sensitivity}}
\end{figure}

All XENON1T tubes are screened before their installation into the detector in order to verify the low radioactivity values of each production batch.  We have screened one batch of  R11410-20 PMTs  and 14 batches of  R11410-21 PMTs with the Gator detector.  Two batches of 4 PMTs were screened with a GeMPI detector. 
The results of both analyses, performed independently, agree with one another within the quoted errors. In the measurements performed with the GeMPI detector, the 4 tubes were placed with the PMT cathode facing the Ge crystal. Due to this configuration, the estimation of the detection efficiency determined by Monte Carlo simulations is affected by the location of the radioactive contamination in the tube. The systematic uncertainty has been evaluated to be up to $\sim$30\,\%. For the measurements with Gator, low-background polytetrafluoroethylene (PTFE) support structures were fabricated to ensure the safe and reproducible handling of all PMTs during the screening process. In Figure~\ref{fig:pmt_holder}  we show a schematic view of 15 tubes in the Ge detector screening chamber. This is the maximum number of tubes that can be measured simultaneously in the PTFE holders, due to space constraints. 
Figure\,\ref{fig:pmt_sensitivity} shows the sensitivity for various isotopes as a function of time, assuming that 15 PMTs are installed in the detector chamber. 

For most relevant isotopes to our study,  a sensitivity below 1\,mBq/PMT can be reached after about  $(4-18)$\,days of measuring time. An example spectrum is shown in Figure\,\ref{fig:pmt_spectra} (bottom), along with the background spectrum of the Ge spectrometer,  including the PTFE PMT holder.
The screening results for $^{238}$U, $^{226}$Ra,  $^{235}$U, $^{228}$Ra, $^{228}$Th, $^{40}$K, $^{60}$Co and $^{110m}$Ag for each batch are shown in Table \ref{tab:screen}. {As the presented upper limits on $^{238}$U are obtained from inspecting the $^{234}$Th and $^{234m}$Pa decays,  we note here that the inferred  $^{238}$U activity from the measured  activity for $^{235}$U is about (8$\pm$1)\,mBq/PMT.

\begin{table*}
\begin{center}
\caption{Screening results for several R11410 PMT batches using a HPGe detector. The number of measured tubes and the measuring time, $t$, in days of each measurement are shown in the first columns. The  $^{137}$Cs upper limit  is in the range $< (0.2-0.3)$\,mBq/PMT for all measured batches.   $^{(1)}$These two sets of 4 PMTs were screened with a different detector. }
\label{tab:screen}
\begin{tabular}{ cc c cccccccc}
\hline
PMT  version & Batch nr. & $t$\,[d]  & \multicolumn{8}{c }{Activity [mBq/PMT]}	\\
(nr. of units)    &               &              &                   &                    &                   &                     &                    &                &                    &                     \\
                       &               &              & $^{238}$U & $^{226}$Ra & $^{235}$U & $^{228}$Ra & $^{228}$Th & $^{40}$K & $^{60}$Co  & $^{110m}$Ag\\
\hline
v-20 (10)           &  0& $15$  & $<18$ & $<0.82$ & $<0.79$ &  $0.9(3)$ & $0.9(2)$ & $12(2)$ & $1.3(2)$  & $<0.21$ \\
v-21 (10)           &  1& $26$  & $<18$ & $0.4(1)$ & $0.5(1)$ & $<1.1$ &    $0.4(1)$ & $12(2)$ & $0.7(1)$  & $0.89(1)$\\
v-21 (16)          &  2 & $15$  & $<16$ & $0.5(1)$ & $0.29(9)$ & $<0.85$ & $<0.61$ & $13(2)$ & $0.79(8)$  &  $1.1(2)$ \\
v-21 (15)          &  3 & $11$  & $<20$ & $<0.82$ & $<0.52$ &   $<1.1$ &  $0.5(2)$ &   $13(2)$ & $0.73(9)$ & $0.51(7)$\\
v-21 (15)          &  4 & $22$  & $<13$ & $0.5(1)$ & $0.35(9)$ & $0.4(1)$ & $0.4(1)$ & $12(2)$ & $0.73(9)$ & $0.21(4)$ \\
v-21 (15)          &  5 & $16$  & $<17$ & $0.6(1)$ & $<0.57$  &  $<0.93$ & $<0.62$ &  $14(2)$ & $0.63(7)$ & $0.22(6)$\\
v-21 (11)          &  6 & $23$  & $<15$ & $0.6(1)$ & $<0.55$  &  $<0.77$ & $0.7(1)$ &  $14(2)$ & $0.71(7)$ & $0.23(4)$\\
v-21 (4)$^{(1)}$ & 6b& $39$  & $ -$ & $0.5(1)$ & $<0.30$  &  $0.3(1)$ & $0.3(1)$ &  $8(1)$ & $0.9(1)$ & $0.24(4)$\\
v-21 (11)          &  7 & $23$  & $<19$ & $1.0(1)$ & $0.4(1)$  &  $<0.77$ & $0.7(1)$ &  $15(2)$ & $1.0(1)$ & $0.22(6)$\\
v-21 (15)          &  8 & $14$  & $<20$ & $0.9(2)$ & $<0.85$  &    $0.7(2)$ & $1.0(2)$ &  $20(3)$ & $1.2(1)$ & $<0.32$\\
v-21 (4)$^{(1)}$ & 8b& $36$  & $ - $ & $0.7(1)$ & $<0.36$  &  $0.3(1)$ & $0.2(1)$ &  $10(1)$ & $1.4(2)$ & $0.17(4)$\\
v-21 (15)          &  9 & $20$  & $<14$ & $0.57(9)$ & $<0.44$  & $<0.79$ & $0.5(1)$ & $13(2)$ & $0.81(8)$ & $0.53(6)$\\
v-21 (15)        &  10 & $26$  & $<15$ & $0.45(7)$ & $<0.44$  & $0.5(1)$ & $0.45(8)$ & $13(2)$ & $0.87(8)$ & $0.60(6)$\\
v-21 (15)        &  11 & $12$  & $<10$ & $0.5(2)$ & $<0.47$  &   $<1.17$ & $0.6(1)$ &  $12(2)$ & $0.77(9)$ & $0.59(7)$\\
v-21 (15)        &  12 & $18$  & $<10$ & $<0.71$ & $<0.45$  &   $0.7(2)$ & $0.7(1)$ &  $11(2)$ & $0.78(8)$ & $0.71(7)$\\
v-21 (15)        &  13 & $34$  & $<10$ & $0.50(6)$ & $0.38(8)$ & $0.6(1)$ & $0.50(7)$ & $12(1)$ & $0.82(7)$ & $0.73(6)$\\
v-21 (15)        &  14 & $21$  & $<16$ & $0.53(8)$ & $<0.41$ & $<0.82$ & $0.5(1)$ & $14(2)$ & $0.81(8)$ & $0.63(6)$\\
\hline
\end{tabular}
\end{center}
\end{table*}
Gamma lines from the anthropogenic isotope $^{110m}$Ag $(\textrm{T}_{1/2}=249.8\,\textrm{d})$ have also been identified in the spectrum. The contamination has been located by two separate measurements with a HPGe detector as being present in the silver brazing between the ceramic stem and the leads and the Kovar flange of the PMT.  The screening results of two samples provided by Hamamatsu with masses of 50\,g and 160\,g, respectively, are consistent with the $^{110m}$Ag activities observed in the individual PMT batches.  

 Comparing the total activities from Table\,\ref{tab:results_U_Th} and Table\,\ref{tab:results_k_co_cs} with the results from the PMT batch screening in Table\,\ref{tab:screen}, we see that these are consistent for all inspected isotopes, with the exception of $^{40}$K and $^{60}$Co. We will discuss these in Section\,\ref{sec:impact}.

In Table~\ref{tab:norm}, we show the screening results normalized by active PMT photocathode area, and compare these with previous versions and other PMT types used in liquid xenon experiments.  We remark that the  R11410-10 activities  obtained by the LUX collaboration \cite{Akerib:2012da}  are slightly lower than those measured by the PandaX \cite{Cao:2014jsa} and the XENON collaborations for the same version of the tube. The R8520 sensor is a 1-inch square tube employed in the XENON100 experiment\,\cite{Aprile:2011dd} and the R8778 is a 2-inch cylindrical tube operated in the LUX detector\,\cite{Akerib:2013tjd}.

 \begin{table*}
\begin{center}
\caption{Screening results normalised per effective photocathode area of the PMT. These data are compared with results from older R11410 versions, as well as from other PMT types employed in liquid xenon detectors. The R8520 PMT is employed in XENON100 \,\cite{Aprile:2011dd} and PandaX  \cite{Cao:2014jsa}, the R8778 PMT is used in the LUX detector\,\cite{Akerib:2013tjd},  and the R11410-10 version in the PandaX experiment\,\cite{Cao:2014jsa}.  The effective photocathode areas are 4.2\,cm$^2$, 15.9\,cm$^2$ and 32.2\,cm$^2$ for the R8520, R8778 and R11410 tubes, respectively. \label{tab:norm}}
\begin{tabular}{lcccccc c }
\hline
PMT  type   & \multicolumn{6}{c }{Normalized activity [mBq/cm$^2$]}	& Ref.\\
                                                     &  $^{238}$U & $^{226}$Ra  & $^{228}$Th   & $^{235}$U  & $^{40}$K & $^{60}$Co& \\
\hline
R11410-21   & $<0.4$ & $0.016(3)$ & $0.012(3)$ & $0.011(3)$ & $0.37(6)$ & $0.023(3)$  &this work \\
R11410-20    & $<0.56$ & $<0.03$  & $0.028(6)$ & $<0.025$&  $0.37(6)$ & $0.040(6)$  & this work\\
R11410-10 & $<3.0$ & $<0.075$ & $<0.08$ &$<0.13$ & $0.4(1)$ & $0.11(2)$ & \cite{Aprile:2011ru}\\
R11410-10 (PandaX) & --& $<0.02$ & $<0.02$ & $0.04(4)$ & $0.5(3)$ & $0.11(1)$ & \cite{Cao:2014jsa}\\
R11410-10 (LUX)  & $<0.19$ & $<0.013$  & $<0.009$ &-- & $<0.26$ & $0.063(6)$ & \cite{Akerib:2012da} \\
R11410 & $1.6(6)$ & $0.19(2)$ & $0.09(2)$ & 0.10(2)& $1.6(3)$ & $0.26(2)$  &\cite{Aprile:2011ru} \\
R8778  (LUX)  & $<1.4$ & $0.59(4)$ & $0.17(2)$ &-- & $4.1(1)$ & $0.160(6)$ &\cite{Akerib:2012da} \\
R8520   & $<0.33$ & $0.029(2)$  & $0.026(2)$ &0.009(2) & $1.8(2)$ & $0.13(1)$  &\cite{Aprile:2011ru} \\
\hline
\end{tabular}
\end{center}
\end{table*}
 The radioactivity per photocathode area of the XE\-NON1T phototube, R11410-21, has been lowered for all isotopes compared with previous versions, and in particular with the one of the R8520 tube employed in the XENON100 experiment.

\section{Discussion and impact on the XENON1T background}\label{sec:impact}

 The contributions of the individual materials to the radioactivity budget of the R11410-21 PMT  are shown in Figure\,\ref{fig::bar_plot}. Each row shows the relative contribution of these materials to the  $^{238}$U, $^{226}$Ra, $^{228}$Ra, $^{228}$Th, $^{40}$K, $^{60}$Co and $^{137}$Cs levels, in those cases when an actual detection was possible. Inferred upper limits on the various isotopes are thus not used to calculate the relative contributions.
 
 \begin{figure*}
  \begin{center}
\includegraphics[width=2.1\columnwidth,keepaspectratio,angle=0]{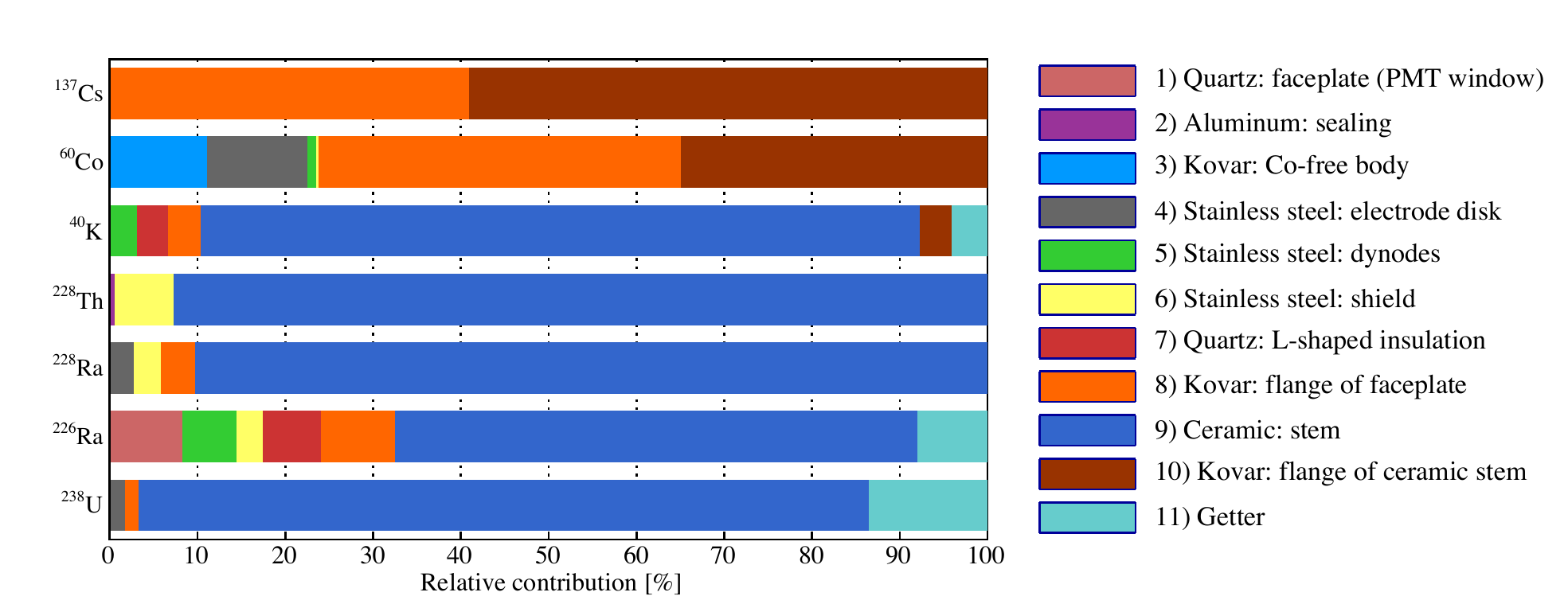}
  \end{center}
 \caption{ Contribution of individual raw materials of the R11410-21 phototube to the total contamination in $^{238}$U, $^{226}$Ra, $^{228}$Ra, $^{228}$Th, $^{40}$K, $^{60}$Co and $^{137}$Cs.  The component numbers correspond to those in Figure \ref{fig:pmt_schematic}  and Table \ref{tab:samples}.   \label{fig::bar_plot}}
\end{figure*}

 The main contribution to the  isotopes in the $^{238}$U and $^{232}$Th  chains clearly comes from the ceramic stem (nr.\,9) of the PMT.
In an attempt to reduce the $^{238}$U and $^{232}$Th activities, sapphire (nr.\,12) was considered as a ceramic replacement. Our measurements show a reduction in $^{238}$U by more than a factor of three, which is, however, counteracted by a $^{226}$Ra level almost three times higher.  The $^{228}$Ra contamination is reduced by less than a factor of two and the $^{228}$Th activity is barely lower than in the case of ceramic (see also Table\,\ref{tab:results_U_Th}).  Monte Carlo simulations using the sapphire screening results show that the neutron-induced background from the PMTs is reduced by less than 10\%. Therefore, such a replacement would not imply a significant improvement in the overall background.

The isotope $^{137}$Cs has been detected only in the two Kovar alloy flanges, namely the flange to the PMT faceplate (nr.\,8) and the one to the stem (nr.\,10) but its overall activity is very low, at the level of 10\,$\mu$Bq/PMT.
In all Kovar alloy  (nr.\,3, 8, 10) and stainless steel (nr.\,4, 5, 6) samples, $^{60}$Co has been detected, the main contribution coming from the Kovar flanges in the PMT faceplate and stem  (nr.\,8 and 10).  }
The total measured $^{60}$Co contamination in the tubes is at the level of 0.8\,mBq/PMT, while the sum of the components yields a range $(0.61\,-\, 0.66)$\,mBq/PMT. The discrepancy could be due to the metallic connections inside the tube that were not screened. 
A further reduction in the $^{60}$Co content could be achieved by substituting the material of the flanges, which is Kovar alloy, with Co-free Kovar.
The main contributors to  $^{40}$K are the getter and the ceramic stem.  However, out of the 13\,mBq of  $^{40}$K detected per tube, only about 1.4\,mBq are located in the ceramic  (nr.\,9). Even taking into account the upper limits on this isotope from all other measured samples, a total upper limit on the $^{40}$K  contamination of $<3.1$\,mBq results. This is significantly lower than the contamination of the final tube. We speculate that the additional $^{40}$K contamination is in the unmeasured cathode material as, in general, bialkali contain potassium. 
This assumption is confirmed by the measurement of the individual getter taken from one phototube, which shows a higher concentration of $^{40}$K compared to the 10 getter samples. The contamination of the getter could have happened during the deposition process of the photocathode material.

The  average radioactive contaminations per PMT have been used in a full GEANT4-based Monte Carlo simulation of XENON1T to quantify the contribution to the background of the experiment. The simulation takes into account a detailed detector geometry, with the 248 R11410-21 PMTs placed at their correct positions in the TPC, above and below the liquid xenon target.
The resulting total electronic-recoil background from gamma interactions in the medium  is  $(0.014\pm0.002)$\,events/y in a central, fiducial volume containing 1~tonne of LXe.  This value is calculated for a dark matter search region of $(2 -12)$\,keV electronic recoil energy and an assumed background discrimination level of 99.75\%.
Although a  contamination from the anthropogenic $^{110m}$Ag has been found in most of the screened PMT batches, the expected contribution to the background of XENON1T by the start of the science run is negligible. This is due to the relatively short half live of 249.8~days of $^{110m}$Ag.

The nuclear-recoil background from neutron interactions  has also been investigated: The estimated rate, taking into account the activity in each  PMT material and the subsequent neutron production rates and spectra, is  ($0.060\pm0.015$)\,events/y  in a 1~tonne  fiducial volume in the  $(5 -50)$\,keV region for nuclear-recoil energies with an  acceptance of 50\%.
 The overall background goal of XENON1T is $<1$\,event for an exposure of 2\,t$\times$y. Thus, we can safely conclude that the PMTs will not limit the sensitivity of the experiment.

\section{Acknowledgments}

We gratefully acknowledge support from NSF, DOE, SNF, UZH, FCT, Region des Pays de la Loire, STCSM, BMBF, MPG, FOM, the Weizmann Institute of Science, EMG, INFN,  and the ITN Invisibles  (Marie Curie Actions, PITN-GA-2011-289442).   We thank the  technical and engineering personnel of Hamamatsu Photonics Co. for the fruitful cooperation in  producing the ultra-pure photomultiplier tube. 
We are grateful to LNGS for hosting and supporting the XENON project.



\end{document}